\documentclass[9pt,letterpaper,aps,pra,notitlepage,twocolumn]{revtex4}
\usepackage{graphicx}
\usepackage{amsmath}
\usepackage{amssymb}
\usepackage{amsmath}
\usepackage{comment}
\usepackage{bm}
\newcommand{\pr}[1]{| #1 \rangle \langle #1|}
\newcommand{\ket}[1]{| #1 \rangle}

\newcommand{\braket}[1]{\langle #1 \rangle}

\newcommand{\eq}[1]{\begin{align}#1\end{align}}

\newcommand{\nn}{\nonumber}

\newcommand{\seq}[1]{\begin{subequations}#1\end{subequations}}
\newcommand{\sseq}[1]{\seq{\eq{#1}}}

\newcommand{\x}{\mathrm{x}}
\newcommand{\diag}{\text{diag}}

\newcommand{\Tor}{\text{Tor}}

\renewcommand{\x}{\hat x}
\newcommand{\p}{\hat p}
\renewcommand{\a}{\hat a}

\newcommand{\V}{\bm{V}}
\newcommand{\tr}{\text{Tr}}
\newcommand{\thr}{threshold}

\renewcommand{\r}{\bm{r}}
\renewcommand{\p}{\hat \rho}

\usepackage{newfloat}
\newcommand{\Q}{\bm{\Sigma}}
\usepackage{algorithmic}
\usepackage{algorithm}
\usepackage{color}
\usepackage{bbm}
\usepackage{pdfpages}

\DeclareMathOperator{\one}{\mathbbm{1}}
\DeclareMathOperator{\id}{\mathbbm{\hat  I}}
\DeclareMathOperator{\Col}{\mathcal{C}}
\newcommand{\beq}{\begin{equation}}
\newcommand{\eeq}{\end{equation}}

\begin{document}

\title{Gaussian Boson Sampling using \thr \ detectors}
\author{Nicol\'as Quesada}
\email{nicolas@xanadu.ai}
\affiliation{Xanadu, 372 Richmond Street W, Toronto, Ontario M5V 1X6, Canada}
\author{Juan Miguel Arrazola}
\email{juanmiguel@xanadu.ai}
\affiliation{Xanadu, 372 Richmond Street W, Toronto, Ontario M5V 1X6, Canada}
\author{Nathan Killoran}
\email{nathan@xanadu.ai}
\affiliation{Xanadu, 372 Richmond Street W, Toronto, Ontario M5V 1X6, Canada}
\begin{abstract}
We study what is arguably the most experimentally appealing Boson Sampling architecture: Gaussian states sampled with threshold detectors. We show that in this setting, the probability of observing a given outcome is related to a matrix function that we name the Torontonian, which plays an analogous role to the permanent or the Hafnian in other models. We also prove that, provided that the probability of observing two or more photons in a single output mode is sufficiently small, our model remains intractable to simulate classically under standard complexity-theoretic conjectures. Finally, we leverage the mathematical simplicity of the model to introduce a physically motivated, exact sampling algorithm for all Boson Sampling models that employ Gaussian states and threshold detectors. 
\end{abstract}

\maketitle

\section{Introduction} Parallel developments in computational complexity theory and quantum optics have raised the possibility of achieving a quantum advantage in sampling problems using non-universal models of quantum computation \cite{harrow2017quantum}. Arguably, the most celebrated of these developments is the Boson Sampling problem \cite{aaronson2011computational} where indistinguishable single photons are sent through a passive linear optics network and then probed using photon counters. Experimental constraints in the generation of indistinguishable single photons have led to the development of new models such as Scattershot Boson Sampling  \cite{lund2014boson,bentivegna2015experimental,latmiral2016towards} and Gaussian Boson Sampling (GBS) \cite{hamilton2017gaussian,kruse2018detailed}, {the latter of which has also been shown to have applications in quantum chemistry \cite{huh2015boson,clements2017experimental,sparrow2018simulating}, optimization \cite{arrazola2018using, arrazola2018quantum}} and graph theory \cite{bradler2018gaussian}. In both of these models, single photons are replaced by squeezed states of light which are amenable to large scale experimental production \cite{yoshikawa2016invited,vernon2018scalable}, but still require photon-number-resolving detectors (PNRs). More recently, new protocols have shifted the experimental complexity back to the state preparation side by replacing single photons with photon-subtracted or photon-added squeezed states that are now probed using heterodyne measurements, which are simpler to perform than photon counting \cite{walschaers2017entanglement,walschaers2017statistical,chabaud2017continuous}. A similar strategy of preparing non-Gaussian states followed by Gaussian measurements was followed in Ref. \cite{lund2017exact}.

Although many of the models listed above have lessened the experimental difficulties of building a Boson Sampler, none of them has looked at what is perhaps the most experimentally accessible configuration: squeezed states undergoing linear operations sampled with \thr \ detectors. These binary outcome detectors measure whether there were 0 photons or 1 or more photons in the field being measured. As opposed to currently available PNRs based on superconducting technology, threshold detectors are inexpensive, commercially available, and can be operated at room temperature \cite{hadfield2009single}. 

In this work, we study the problem of sampling Gaussian states using threshold detectors. In the same way that the probability distribution of regular Boson Sampling is related to the permanent, and in GBS to the Hafnian, when sampling Gaussian states with threshold detectors the output distribution is related to a matrix function that we name the Torontonian. The Torontonian can be interpreted as an infinite sum of Hafnians. We also prove that, if in GBS the probability of observing two or more photons in the same output mode is sufficiently small, the model remains hard to simulate classically even when employing threshold detectors.  {We also propose a new physically motivated exact classical sampling algorithm which can be used for all the Boson Sampling models mentioned above when employing threshold detectors.}  This constitutes the first explicit example of a classical sampling algorithm for Boson Sampling based on Gaussian states, with a running time whose only source of exponential growth is the number of non-Gaussian events (clicks) in the sampling. A recent benchmarking of the algorithm presented here has been done in Ref. \cite{gupt2018classical} using the Titan supercomputer from Oak Ridge National Laboratory where it was found that a 20 click sample from an 800 mode system can be obtained in about two hours using 240000 CPUs.

\section{Gaussian states} Gaussian states form an experimentally accessible set of states that can be efficiently described in the symplectic formalism in terms of covariance matrices and mean vectors \cite{weedbrook2012gaussian,serafini2017quantum}. In this description, we arrange the canonical operators of the $\ell$ modes of interest in a vector 
$\bm{\hat r} = (\x_1,\hat p_1,\ldots,\x_\ell,\hat p_\ell)^T$.
Gaussian states $\rho(\bm{V}, \bar{ \bm{r}})$ have the special property that they are completely characterized by a vector of means $\bar{ \bm{r}} = \langle \hat{\bm{r}} \rangle_{\rho} = \text{Tr}\left(\hat{ \bm{r}}  \rho \right) $  and a covariance matrix $\bm{V}_{ij} = \frac{1}{2} \langle \Delta \hat r_i \Delta \hat r_j+ \Delta \hat r_i \Delta \hat r_j \rangle_{\rho}$ with $\Delta \hat{\bm{ r}}=\hat{ \bm{r}}-\bar{ \bm{r}}$. 
{For the {multimode} vacuum state $\bm{V} = \one_{2 \ell}$ (using the convention $\hbar =2$) and $\bm{\bar r} = 0$}.
It will also be useful to employ the $Q$ function of the Gaussian state $\rho$, defined as $Q(\bm{\alpha}) = \braket{\bm{\alpha} | \rho | \bm{\alpha}}/\pi^\ell =  \exp\left(-\Delta \bm{\alpha} \ \Q^{-1} \ \Delta \bm{\alpha}^\dagger /2 \right)/(\pi^\ell \sqrt{\det(\Q)})$ with $\Delta \bm{\alpha}=\bm{\alpha}-\bm{\bar \alpha} $. The covariance matrix of the Gaussian $Q$ function is
\beq
\Q = \frac{1}{4} \left(\bm{B C} \right) \bm{V} \left( \bm{BC} \right)^\dagger + \frac{1}{2} \one_{2 \ell},
\eeq
is the covariance matrix of the complex amplitudes $\bm{\alpha} = \bm{x}+i \bm{p}$ and their complex conjugates $\bm{\alpha}^* = \bm{x}-i \bm{p}$. Here $\bm{B}$ is the permutation matrix that takes the vector $\bm{\hat r}$ to the $xp-$ordering $(\hat x_1,\ldots,\hat x_\ell, \hat p_1,\ldots,\hat p_\ell) = (\bm{\hat  x},\bm{\hat p})$, and $\bm{C} =  \left[ \begin{smallmatrix}
\one_{\ell} & i \one_{\ell} \\
\one_{\ell} & - i \one_{\ell}  
\end{smallmatrix} \right]$. Finally, note that both $\Q$ and its inverse have the following block structure
\eq{\label{block}
	\Q=\begin{bmatrix}
	\bm{W} & \bm{Y}^* \\
	\bm{Y} & \bm{W}^* \\
	\end{bmatrix},
}
where  $\bm{W} = \bm{W} ^\dagger  \in \mathbb{C}^{\ell \times \ell}$ is Hermitian  and $\bm{Y} = \bm{Y} ^T \in \mathbb{C}^{\ell \times \ell}$ is symmetric.
{As in previous works, we focus on the case of zero displacement, $\bm{\bar{\alpha}}=\bm{\bar{x}}=\bm{\bar{p}}=0$.} 

\section{Click probabilities and Torontonians} {It is well known that the combination of Gaussian states and Gaussian measurements can be efficiently simulated on a classical computer \cite{bartlett2002universal,bartlett2002efficient,serafini2017quantum}.}
An experimentally accessible non-Gaussian measurement is the one performed by \thr \ detectors \cite{lita2008counting,marsili2013detecting,fiuravsek2005conditional}. These detectors perform a measurement defined by the POVM elements
\beq\label{effects}
\hat \Pi^{(n)}_0 = \pr{0_n}, \quad \hat \Pi^{(n)}_1 = \id^{(n)}-\hat \Pi^{(n)}_0,
\eeq
where $\id^{(n)}$ is the identity operator in the Hilbert space of mode $n$ and $\ket{0_n}$ is the vacuum state of mode $\a_n$. The outcome $\hat \Pi^{(n)}_1$ corresponds to a click in the detector and $\hat \Pi^{(n)}_0$ to no click.

When using photon counting to measure an $\ell$-mode Gaussian state, we denote a particular outcome 
(with $N$ total photons) 
by a multiset $S = \{ i_1,i_2,\ldots,i_{N}\}$ specifying the modes where photons were detected. 
The multiplicity of mode-index $k$, denoted $s_k$, is the number of photons that were detected in that mode, with $\sum_{k=1}^{\ell} s_k=N$.
The probability of the outcome $S$ is \cite{hamilton2017gaussian,kruse2018detailed}
\eq{\label{Eq:psHaf}
p(S)=\frac{\text{Haf}[\bm{X}\bm{O}_{(S)}]}{\sqrt{\det(\Q)}s_{1}!\dots s_{\ell}!},
}
where $\bm{X} =  \left[ \begin{smallmatrix}
	0 &  \one \\
	\one & 0  
\end{smallmatrix} \right]$,
$\bm{O}_{(S)} = \one  - (\Q^{-1})_{(S)}$ and
$\bm{A}_{(S)}\in\mathbb{C}^{2N\times 2N}$ is the matrix formed by indexing elements within each block of $A$ according to the multiset $S$. More precisely, if mode-index $k$ has multiplicity $s_k$, the corresponding row and column of $\bm{A}$ is repeated (or dropped when $s_k=0$) from each block when forming $\bm{A}_{(S)}$.
For example if one has three modes and writes
\eq{
\bm{A} &= 	\left[
\begin{array}{cc}
\bm{W} & \bm{Y}^* \\
\bm{Y} & \bm{W}^*\\
\end{array}
\right],\\
\bm{W} &= \left[
\begin{array}{ccc}
	W_{1,1} & W_{1,2} & W_{1,3} \\
	W_{2,1} & W_{2,2} & W_{2,3} \\
	W_{3,1} & W_{3,2} & W_{3,3}
\end{array}
\right], 
\bm{Y} = \left[
\begin{array}{ccc}
	Y_{1,1} & Y_{1,2} & Y_{1,3} \\
	Y_{2,1} & Y_{2,2} & Y_{2,3} \\
	Y_{3,1} & Y_{3,2} & Y_{3,3}
\end{array}
\right] \nonumber
}
and has $s_1=3,s_2=0,s_3=1$ then
\eq{
	\bm{A}_{(S)} &= 	\left[
	\begin{array}{cc}
		\bm{W}_{(s)} & \bm{Y}^*_{(s)} \\
		\bm{Y}_{(s)} & \bm{W}^*_{(s)}\\
	\end{array}
	\right],\\
\bm{W}_{(s)} &= \left[
\begin{array}{cccc}	
W_{11} &   W_{11} & W_{11} & W_{13}    \\
W_{11} &  W_{11}  & W_{11}  & W_{13}   \\
W_{11} & W_{11}  & W_{11} & W_{13}  \\
W_{31} & W_{31} & W_{31}  & W_{33} 
\end{array}
\right],\\
\bm{Y}_{(s)} &= \left[
\begin{array}{cccc}	
	Y_{11} &   Y_{11} & Y_{11} & Y_{13}    \\
	Y_{11} &  Y_{11}  & Y_{11}  & Y_{13}   \\
	Y_{11} & Y_{11}  & Y_{11} & Y_{13}  \\
	Y_{31} & Y_{31} & Y_{31}  & Y_{33} 
\end{array}
\right].
}

The same notation can be employed when using threshold detectors, in which case the elements of $S$ correspond to the modes where a click was observed, and no element has multiplicity greater than one. The probability of observing an outcome $S$ is given by
\eq{\label{ps}
p(S) = \pi^\ell \int  \prod_{i \in S} d^2 \alpha_i P^{(i)}_{1}(\alpha_i) 
 \prod_{k \notin S} d^2 \alpha_k P^{(k)}_{0}(\alpha_k) 
 Q(\bm{\alpha}),
}	
where the $P$ functions of the POVM elements in Eq. (\ref{effects}) are
\sseq{\label{vac0P}
P^{(n)}_0(\alpha_n) &= \delta^{(2)}(\alpha_n) = \delta(\alpha_n)\delta(\alpha_n^*), \\
P^{(n)}_1(\alpha_n) &= \frac{1}{\pi} -P^{(n)}_0(\alpha_n).
}
{By performing a straightforward yet lengthy calculation (see Appendix \ref{derivation} for details), it is possible to show that the probability of an outcome $S$ is given by}
\eq{\label{Eq:psTor}
p(S)=\frac{\text{Tor}[\bm{O}_{(S)}]}{\sqrt{\det(\Q)}},
}
where 
\eq{\label{Eq:Torontonian}
\text{Tor}(\bm{A}) = \sum_{Z \in P([N])} (-1)^{|Z|}  \frac{1}{\sqrt{\det(\one - \bm{A}_{(Z)})}}
}
is the \emph{Torontonian} of a matrix $\bm{A} \in \mathbb{C}^{2N \times 2N}$ with a block structure as in Eq. (\ref{block}). Here $P([N])$ is the power set (the set of all subsets) of $[N]:=\{1,2,\ldots,N\}$. 
Note that a direct calculation of the Torontonian according to Eq.~\eqref{Eq:Torontonian} requires the computation of $2^N$ determinants. When the determinants are calculated using standard algorithms based on Cholesky decompositions, this leads to a complexity of $O(N^32^N)$ for a direct calculation of the Torontonian, which is equivalent to the state-of-the-art for computing Hafnians \cite{bjorklund2018faster}.
 
The probability of a certain click pattern $S$ obtained with \thr \ detectors can also be computed by summing all the corresponding probabilities of that event when using PNRs. 
{Given a threshold click pattern $S=\{i_1,i_2,\ldots,i_N\}$, let $\mathcal{C}_S$ be the set of all outcomes where photons are observed {only in the modes $i_k\in S$ and there is at least one mode with multiplicity $s_k>1$, i.e., with a collision in that mode.} From Eqs.~\eqref{Eq:psHaf} and \eqref{Eq:psTor}, it holds that
\begin{align}
\text{Tor}[\bm{O}_{(S)}]=\text{Haf}[\bm{X O}_{(S)}]+\sum_{S'\in\mathcal{C}_S}\frac{\text{Haf}[\bm{X O}_{(S')}]}{s'_{1}!\cdots s'_{\ell}!}\label{Eq:TorHaf}.
\end{align}
This equation} suggests that the Torontonian is a kind of generating function for all the PNR click statistics, which are all proportional to Hafnians. As shown in detail in Appendices \ref{app2} and \ref{app3} , this link can be formalized by using recently developed algorithms for the calculation of Hafnians \cite{bjorklund2018faster}, leading to the expression
\eq{
	\label{hafmgf}
	\text{Haf}(\bm{X} \bm{O}) 
	&=\frac{1}{\ell!} \left. \frac{d^\ell}{d\eta^\ell} \text{Tor}(\eta \bm{O}) \right|_{\eta=0},
}
where the matrix $\bm{O}$ has size $2\ell \times 2 \ell$.

\section{Complexity of threshold GBS} When sampling from a Gaussian state in a regime where there is a very small probability of observing two or more photons in the same output mode, the use of threshold detectors should not significantly affect the properties of the underlying distribution. Formally, let $p(S)$ be the probability of observing an output pattern $S$ when sampling from a state using PNRs and let $p'(S)$ be the probability when using threshold detectors. 
{We define $p'(S)=0$ for any $S$ whose elements have multiplicity greater than one, since those patterns will not appear in a threshold experiment.}
Define the set of all collision outputs $\mathcal{C}$ as the set of PNR outputs where two or more photons are observed in at least one mode. The probability of observing a collision when sampling from $p(S)$ is then $\varepsilon := \sum_{S\in \mathcal{C}} p(S)$. As shown in the Appendix \ref{app5}, the distance between these two distributions satisfies
\beq\label{Eq: trace dist}
\|p(S)-p'(S)\|_1=\varepsilon,
\eeq
confirming that the distributions are difficult to distinguish when $\varepsilon$ is very small. {Let $q(N)$ be the probability of detecting $N$ photons in an $\ell$-mode state obtained by sending a Gaussian state through a linear interferometer characterized by a unitary $\mathcal{U}$.} It then holds that 
\beq
\mathbb{E}_{\mathcal{U}}[\varepsilon]=\frac{8}{\ell}\mathbb{E}_q[N^2],
\eeq
where the first expectation is taken from the Haar measure and the second expectation is over $q(N)$ (see Appendix \ref{app5} for details). By choosing $\ell=O(\mathbb{E}_q[N^2])$ it is thus possible to set the collision probability to be any fixed small constant. 

In Ref.~\cite{hamilton2017gaussian}, it was shown that if the Hafnian-of-Gaussians conjecture and the Hafnian-anti-concentration conjecture are true, for any fixed $\epsilon>0$ the existence of a polynomial-time classical algorithm that samples from a distribution that is $\epsilon$-close in total variation distance to the output distribution $p(S)$ would imply the collapse of the polynomial hierarchy to the third level. {The setting of small collision probability, used in all previous variants of Boson Sampling, is also the }{regime where the Hafnian-of-Gaussians conjecture applies \cite{hamilton2017gaussian}}. From the above results, it is straightforward to extend this claim to threshold GBS.

Assume that there exists a polynomial-time classical algorithm that samples from a distribution {$\pi(S)$} such that  $\|p'(S)-\pi(S)\|_1=\varepsilon'$ for some $\varepsilon'>0$. From Eq. \eqref{Eq: trace dist} and the triangle inequality it holds that
\begin{align*}
\|p(S)-\pi(S)\|_1&=\|p(S)-p'(S)+p'(S)-\pi(S)\|_1\\
&\leq \|p(S)-p'(S)\|_1 + \|p'(S)-\pi(S)\|_1\\
&= \varepsilon+\varepsilon'.
\end{align*}
Therefore, by setting $\epsilon = \varepsilon+\varepsilon'$, we conclude that the existence of a polynomial-time classical sampling algorithm for threshold GBS also implies a polynomial-time algorithm for GBS with PNRs and consequently a collapse of the polynomial hierarchy to the third level, provided that the Hafnian-of-Gaussians conjecture and the Hafnian-anti-concentration conjecture are true.

\section{Sampling algorithm}  \label{sec:sampling}
\begin{figure}
	\includegraphics[width=0.45\textwidth]{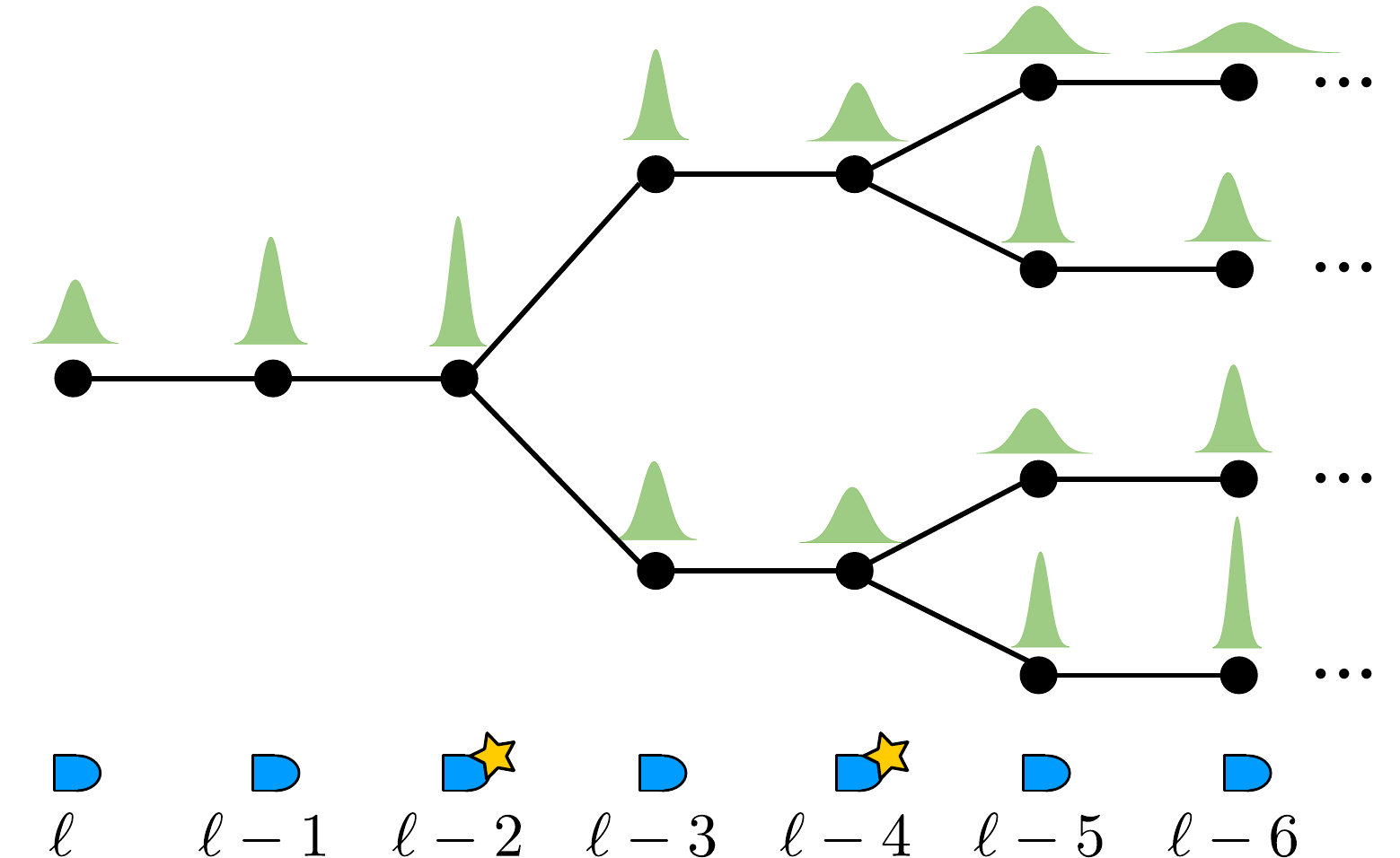}
	\caption{\label{tree} {Schematic illustration of the sampling algorithm. Starting from an $\ell$-mode Gaussian state, we iteratively apply the update rule of Algorithm \ref{Qsampling} for each mode, causing the conditional state of the remaining modes to change, as illustrated by the varying Gaussian curves. In this example, a click occurs in modes $\ell-2$ and $\ell-4$. Each click causes a doubling of the number of Gaussian states in the linear combination $\rho_{\ell'} = \sum_{k} a_k \rho_{\ell'\,k}$ that describes the state of the remaining modes. The complexity of the algorithm grows exponentially in the number of clicks.} }
\end{figure}

{Alongside the development of various Boson Sampling models, there has also been progress in developing classical methods for simulating the original Boson Sampling model of Ref.~\citep{aaronson2011computational}, where approximate Markov chain  \cite{neville2017classical} and exact sampling algorithms 
\cite{clifford2018classical} represent the state of the art. In this section, we show that the appeal of threshold detectors is not only experimental: their action on Gaussian states also has a simple mathematical formulation. We leverage this fact to describe an exact sampling algorithm for threshold GBS.} {This algorithm, shown schematically in Fig. \ref{tree}, can also be adapted to other Boson Sampling settings}.

\begin{figure*}[!ht]
	\includegraphics[width=\textwidth]{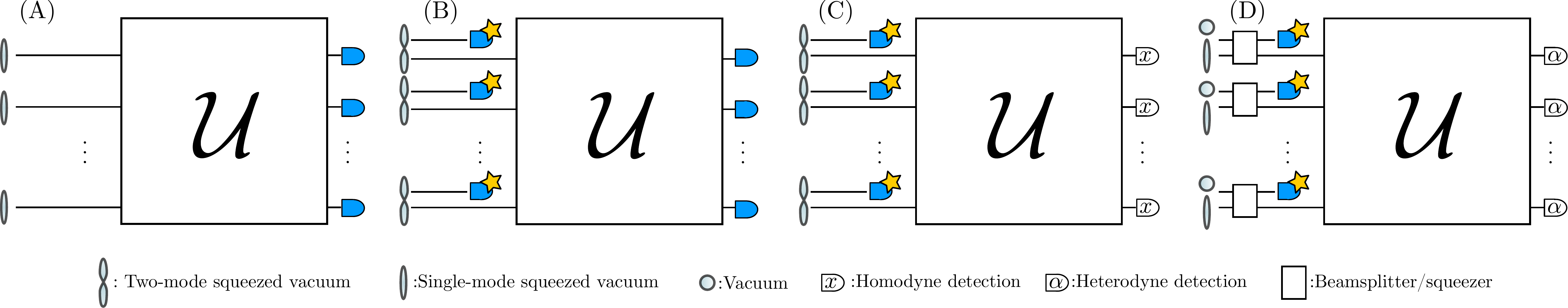}
	\caption{\label{comp} Boson Sampling models using threshold detectors. (A) Gaussian Boson Sampling, where single-mode squeezed states passing through a linear optical interferometer are probed with threshold detectors. (B) Scattershot Boson Sampling, where single photons are prepared by heralding on a click in a \thr \ detector. (C) The model of Ref. \cite{lund2017exact} where heralded single photons are measured using homodyne detection. (D) The protocol of Ref. \cite{chabaud2017continuous} where photon-added or -subtracted states are sent into a linear optical network and then measured using heterodyne detection.}
\end{figure*}

Consider an $\ell$-mode Gaussian state $\rho_{\ell}(\bm{V},\bm{\bar{r}})$ and perform a measurement on the $\ell$-th mode using the POVM of Eq. \eqref{effects}. If no click is observed, {since the operator $\hat{\Pi}_0$ is Gaussian}, the state of the remaining $\ell-1$ modes is also a Gaussian state $\rho_{\ell-1}(\bm{V}_A',\bm{\bar{r}}_A')$ with updated covariance matrix $\bm{V}_A'$ and displacement vector $\bm{\bar{r}}_A'$. This occurs with probability $p=\text{Tr}(\rho_{\ell}\Pi_0^{(\ell)})$. If a click is observed, the conditional state of the remaining modes is a linear combination of Gaussian states given by
\eq{
	\label{singleph}
\rho_{\ell-1} = \frac{\text{Tr}_{\ell}\left(\rho_{\ell} \hat \Pi_1^{(n)}\right)}{1-p}	 = \frac{ \rho_{\ell-1}(\bm{V}_A,\bm{\bar r}_{A})  - p \ \rho_{\ell-1} (\bm{V}_A',\bm{\bar r}_{A}')
	}{1-p}.
}
{Note that in this case $\rho_{\ell-1}$ is a non-Gaussian state.} 
This fact forms the basis of the sampling algorithm: the initial state $\rho_{\ell}$  is iterated through one mode at a time, updating the conditional state using Eq. \eqref{singleph} every time a click is detected, while keeping track of the modes where clicks have been observed. {Suppose that after the $k$th step, corresponding to mode $\ell' = \ell-k$, we have recorded $m$ clicks. Then the tree in Fig.~\ref{tree} has $2^m$ branches at that step, and}
the conditional state can be written as a linear combination of Gaussian states of the form
\beq\label{lc}
\rho_{\ell'} = \sum_{k=1}^{2^m} a_k \rho_{\ell'\,k},
\eeq 
where 
the coefficients {$a_k$ are not all positive in general.} 
The explicit update rule is described in pseudocode in Algorithm \ref{Qsampling}. 
{After iterating through all $\ell$ modes, suppose we have observed $N$ clicks. Let $c_j$ denote the number of steps between clicks $(j-1)$ and $j$, i.e., the number of steps between consecutive branching events. Then a total of $\sum_{j=1}^{N} c_j 2^j=O(2^N)$ probabilities and updates must be computed. In calculating them, the dominant term is the matrix multiplication of Eq.~\eqref{Eq:MatrixUpdate} which requires $O(\ell^2)$ steps, leading to a total complexity of $O(\ell^2 2^N)$.}

\begin{algorithm}[t!]
\caption{Update rule. \label{Qsampling}}
	\begin{algorithmic}
		\STATE{{\bf Input:} $\ell$-mode state: $
			\rho_{\ell} = {\sum_{k=1}^{2^m}} a_k \rho_{\ell,k}(\bm{V}_k,\bm{\bar r}_k)
			$
		}
		\FOR{$1 \leq k \leq N$} 
		\STATE\eq{
			\V_k &{\to} \begin{bmatrix}
				\noindent	\V_{A,k} & \bm{V}_{AB,k} \\
				\bm{V}_{AB,k}^T & \bm{V}_{B,k}
			\end{bmatrix}, \ \bm{\bar r}_k \to \begin{bmatrix} \bm{\bar r}_{A,k} \\ \bm{\bar r}_{B,k} \end{bmatrix}\\
			\bm{V}_{A,k}' &\to \bm{V}_{A,k} - \bm{V}_{AB,k} (\bm{V}_{B,k}+\one_2 )^{-1} \bm{V}_{AB,k}^T,\label{Eq:MatrixUpdate}\\
			\bm{\bar r}_{A,k}'&\to\bm{\bar r}_{A,k}- \bm{V}_{AB,k} (\bm{V}_{B,k}+\one_2)^{-1} \bm{\bar r}_{B,k}
		}
{\STATE\COMMENT{$\bm{V}_{A,k}$ is a $2(\ell-1)\times 2(\ell-1)$ matrix describing modes 1 to $\ell-1$, $\bm{V}_{AB,k}$ is a $2(\ell-1) \times 2$ matrix describing the correlations between modes 1 to $\ell-1,k$ and mode $\ell$ and $\bm{V}_{B,k}$ is a $2 \times 2$ matrix describing mode $\ell$.}}
		\ENDFOR	
		\STATE{Calculate click probability:
			\eq{
				p = \sum_{k=1}^{2^m}  a_k q_k \text{ with } q_k=\frac{2  e^{-\bm{\bar r}_{B,k}^T (\bm{V}_{B,k}+\one_2)^{-1} \bm{\bar r}_{B,k} }}{\sqrt{\text{det}\left( \bm{V}_{B,k} +\one_2 \right)}}
			}
		}
		\STATE Flip a coin with bias $p$
		\IF{click} 
		\STATE\eq{
			\rho_{\ell-1} \to & \sum_{k=1}^{2^m} a_k \frac{\rho_{\ell-1,k}(\bm{V}_{A,k},\bm{\bar r_{A,k}}) -q_k \rho_{\ell-1,k}(\bm{V}_{A,k}',\bm{\bar r}_{A,k}') }{1-p}  
		}
		\ELSE{}
		\STATE\eq{\rho_{\ell-1} \to \sum_{k=1}^{2^m} \left(\frac{a_k q_k}{p}  \right)  \rho_{\ell-1,k}(\bm{V}_{A,k}',\bm{\bar r_{A,k}}')
		}
		\ENDIF
		\STATE{{\bf Output:} $(\ell-1)$-mode state: $\rho_{\ell-1}$}
	\end{algorithmic}
\end{algorithm}

Any passive or active linear optical operation on states that are linear combinations of Gaussian states (as in Eq. \eqref{lc}) can be described by transforming the covariance matrices $\bm{V}_k$ and vectors of means $\bm{\bar r}_k$ of each individual Gaussian state. This includes unitary operations like phase shifts, beamsplitters, and squeezing, as well as {non{-}unitary} operations like loss and linear amplification. Furthermore, at the price of dealing with probability density functions that are linear combinations of two-dimensional Gaussians, we can also simulate single-mode homodyne and heterodyne measurements. When the states in Eq. (\ref{lc}) are probed with PNRs, it is also possible to show that the probabilities of detection are weighted sums of Hafnians (see the Appendix \ref{app3} for details). 

The sampling algorithm presented above can be used to study many different types of Boson Sampling problems. In Fig. \ref{comp}, we summarize the relationship between these models and threshold GBS. As discussed above, $O(\ell^2 2^N)$ operations are required to generate a sample with $N$ clicks from threshold GBS. For Boson Sampling using heralded single photons, as in Fig.~\ref{comp}(B), $N$ clicks are needed to herald $N$ single photons, which are followed by $N$ detections, giving a complexity of $O(\ell^2 2^{2N})$. This scaling does not change for scattershot Boson Sampling where the heralding is moved after the interferometer. As shown in Fig.~\ref{comp}(C), for $N$ heralded single photons undergoing homodyne detection \cite{lund2017exact},  $O(\ell^2 2^N)$ operations are needed to simulate homodyne detection in a single mode, leading to $O(\ell^3 2^N)$ complexity across all $\ell$ modes. The same scaling holds when replacing heralding with photon addition/subtraction and heterodyne  measurements \cite{chabaud2017continuous} as shown in Fig.~\ref{comp}(D). The discussion of how to implement heterodyne and homodyne measurements in states that are linear combinations of Gaussian states can be found in Appendix \ref{app4}.

In all the models considered in the previous paragraph, our sampling algorithm has a scaling that grows exponentially only on the number of clicks obtained. This is similar to the best known classical algorithms \cite{clifford2018classical,neville2017classical} for Boson Sampling in which the complexity of generating a sample scales like $2^n$ where $n$ is the number of photon clicks in the sample.

\section{Conclusion} The experimental appeal of threshold detectors in Boson Sampling is clear: they are standard, inexpensive equipment that can be operated at room temperature. In this work, we have shown that the use of threshold detectors also gives rise to a GBS model that is both mathematically elegant and intractable to simulate classically. At the core of this model is a matrix function -- the Torontonian -- that determines the probability distribution of measurement outcomes, analogously to the role of the permanent and the Hafnian in other variants of Boson Sampling. Our results also lead to a physically-motivated, exact sampling algorithm for all models of Boson Sampling that can be approximated as Gaussian states being measured using threshold detectors. Because the algorithm is exact, it provides an ideal tool for benchmarking near-term Gaussian Boson Sampling devices. Overall, explicitly incorporating threshold detectors may lead to further advances in both the theory and experiment of Boson Sampling.

\section*{Acknowledgments} The authors thank B. Gupt, C. Weedbrook and A. Ignjatovic for valuable discussions.

\bibliographystyle{apsrev}
\bibliography{Torontonian}

\begin{thebibliography}{37}
\expandafter\ifx\csname natexlab\endcsname\relax\def\natexlab#1{#1}\fi
\expandafter\ifx\csname bibnamefont\endcsname\relax
  \def\bibnamefont#1{#1}\fi
\expandafter\ifx\csname bibfnamefont\endcsname\relax
  \def\bibfnamefont#1{#1}\fi
\expandafter\ifx\csname citenamefont\endcsname\relax
  \def\citenamefont#1{#1}\fi
\expandafter\ifx\csname url\endcsname\relax
  \def\url#1{\texttt{#1}}\fi
\expandafter\ifx\csname urlprefix\endcsname\relax\def\urlprefix{URL }\fi
\providecommand{\bibinfo}[2]{#2}
\providecommand{\eprint}[2][]{\url{#2}}

\bibitem[{\citenamefont{Harrow and Montanaro}(2017)}]{harrow2017quantum}
\bibinfo{author}{\bibfnamefont{A.~W.} \bibnamefont{Harrow}} \bibnamefont{and}
  \bibinfo{author}{\bibfnamefont{A.}~\bibnamefont{Montanaro}},
  \bibinfo{journal}{Nature} \textbf{\bibinfo{volume}{549}},
  \bibinfo{pages}{203} (\bibinfo{year}{2017}).

\bibitem[{\citenamefont{Aaronson and
  Arkhipov}(2011)}]{aaronson2011computational}
\bibinfo{author}{\bibfnamefont{S.}~\bibnamefont{Aaronson}} \bibnamefont{and}
  \bibinfo{author}{\bibfnamefont{A.}~\bibnamefont{Arkhipov}}, in
  \emph{\bibinfo{booktitle}{Proceedings of the forty-third annual ACM symposium
  on Theory of computing}} (\bibinfo{organization}{ACM}, \bibinfo{year}{2011}),
  pp. \bibinfo{pages}{333--342}.

\bibitem[{\citenamefont{Lund et~al.}(2014)\citenamefont{Lund, Laing,
  Rahimi-Keshari, Rudolph, O’Brien, and Ralph}}]{lund2014boson}
\bibinfo{author}{\bibfnamefont{A.}~\bibnamefont{Lund}},
  \bibinfo{author}{\bibfnamefont{A.}~\bibnamefont{Laing}},
  \bibinfo{author}{\bibfnamefont{S.}~\bibnamefont{Rahimi-Keshari}},
  \bibinfo{author}{\bibfnamefont{T.}~\bibnamefont{Rudolph}},
  \bibinfo{author}{\bibfnamefont{J.~L.} \bibnamefont{O’Brien}},
  \bibnamefont{and} \bibinfo{author}{\bibfnamefont{T.}~\bibnamefont{Ralph}},
  \bibinfo{journal}{Physical Review Letters} \textbf{\bibinfo{volume}{113}},
  \bibinfo{pages}{100502} (\bibinfo{year}{2014}).

\bibitem[{\citenamefont{Bentivegna et~al.}(2015)\citenamefont{Bentivegna,
  Spagnolo, Vitelli, Flamini, Viggianiello, Latmiral, Mataloni, Brod,
  Galv{\~a}o, Crespi et~al.}}]{bentivegna2015experimental}
\bibinfo{author}{\bibfnamefont{M.}~\bibnamefont{Bentivegna}},
  \bibinfo{author}{\bibfnamefont{N.}~\bibnamefont{Spagnolo}},
  \bibinfo{author}{\bibfnamefont{C.}~\bibnamefont{Vitelli}},
  \bibinfo{author}{\bibfnamefont{F.}~\bibnamefont{Flamini}},
  \bibinfo{author}{\bibfnamefont{N.}~\bibnamefont{Viggianiello}},
  \bibinfo{author}{\bibfnamefont{L.}~\bibnamefont{Latmiral}},
  \bibinfo{author}{\bibfnamefont{P.}~\bibnamefont{Mataloni}},
  \bibinfo{author}{\bibfnamefont{D.~J.} \bibnamefont{Brod}},
  \bibinfo{author}{\bibfnamefont{E.~F.} \bibnamefont{Galv{\~a}o}},
  \bibinfo{author}{\bibfnamefont{A.}~\bibnamefont{Crespi}},
  \bibnamefont{et~al.}, \bibinfo{journal}{Science {A}dvances}
  \textbf{\bibinfo{volume}{1}}, \bibinfo{pages}{e1400255}
  (\bibinfo{year}{2015}).

\bibitem[{\citenamefont{Latmiral et~al.}(2016)\citenamefont{Latmiral, Spagnolo,
  and Sciarrino}}]{latmiral2016towards}
\bibinfo{author}{\bibfnamefont{L.}~\bibnamefont{Latmiral}},
  \bibinfo{author}{\bibfnamefont{N.}~\bibnamefont{Spagnolo}}, \bibnamefont{and}
  \bibinfo{author}{\bibfnamefont{F.}~\bibnamefont{Sciarrino}},
  \bibinfo{journal}{New Journal of Physics} \textbf{\bibinfo{volume}{18}},
  \bibinfo{pages}{113008} (\bibinfo{year}{2016}).

\bibitem[{\citenamefont{Hamilton et~al.}(2017)\citenamefont{Hamilton, Kruse,
  Sansoni, Barkhofen, Silberhorn, and Jex}}]{hamilton2017gaussian}
\bibinfo{author}{\bibfnamefont{C.~S.} \bibnamefont{Hamilton}},
  \bibinfo{author}{\bibfnamefont{R.}~\bibnamefont{Kruse}},
  \bibinfo{author}{\bibfnamefont{L.}~\bibnamefont{Sansoni}},
  \bibinfo{author}{\bibfnamefont{S.}~\bibnamefont{Barkhofen}},
  \bibinfo{author}{\bibfnamefont{C.}~\bibnamefont{Silberhorn}},
  \bibnamefont{and} \bibinfo{author}{\bibfnamefont{I.}~\bibnamefont{Jex}},
  \bibinfo{journal}{Physical Review Letters} \textbf{\bibinfo{volume}{119}},
  \bibinfo{pages}{170501} (\bibinfo{year}{2017}).

\bibitem[{\citenamefont{Kruse et~al.}(2018)\citenamefont{Kruse, Hamilton,
  Sansoni, Barkhofen, Silberhorn, and Jex}}]{kruse2018detailed}
\bibinfo{author}{\bibfnamefont{R.}~\bibnamefont{Kruse}},
  \bibinfo{author}{\bibfnamefont{C.~S.} \bibnamefont{Hamilton}},
  \bibinfo{author}{\bibfnamefont{L.}~\bibnamefont{Sansoni}},
  \bibinfo{author}{\bibfnamefont{S.}~\bibnamefont{Barkhofen}},
  \bibinfo{author}{\bibfnamefont{C.}~\bibnamefont{Silberhorn}},
  \bibnamefont{and} \bibinfo{author}{\bibfnamefont{I.}~\bibnamefont{Jex}},
  \bibinfo{journal}{arXiv:1801.07488}  (\bibinfo{year}{2018}).

\bibitem[{\citenamefont{Huh et~al.}(2015)\citenamefont{Huh, Guerreschi,
  Peropadre, McClean, and Aspuru-Guzik}}]{huh2015boson}
\bibinfo{author}{\bibfnamefont{J.}~\bibnamefont{Huh}},
  \bibinfo{author}{\bibfnamefont{G.~G.} \bibnamefont{Guerreschi}},
  \bibinfo{author}{\bibfnamefont{B.}~\bibnamefont{Peropadre}},
  \bibinfo{author}{\bibfnamefont{J.~R.} \bibnamefont{McClean}},
  \bibnamefont{and}
  \bibinfo{author}{\bibfnamefont{A.}~\bibnamefont{Aspuru-Guzik}},
  \bibinfo{journal}{Nature Photonics} \textbf{\bibinfo{volume}{9}},
  \bibinfo{pages}{615} (\bibinfo{year}{2015}).

\bibitem[{\citenamefont{Clements et~al.}(2017)\citenamefont{Clements, Renema,
  Eckstein, Valido, Lita, Gerrits, Nam, Kolthammer, Huh, and
  Walmsley}}]{clements2017experimental}
\bibinfo{author}{\bibfnamefont{W.~R.} \bibnamefont{Clements}},
  \bibinfo{author}{\bibfnamefont{J.~J.} \bibnamefont{Renema}},
  \bibinfo{author}{\bibfnamefont{A.}~\bibnamefont{Eckstein}},
  \bibinfo{author}{\bibfnamefont{A.~A.} \bibnamefont{Valido}},
  \bibinfo{author}{\bibfnamefont{A.}~\bibnamefont{Lita}},
  \bibinfo{author}{\bibfnamefont{T.}~\bibnamefont{Gerrits}},
  \bibinfo{author}{\bibfnamefont{S.~W.} \bibnamefont{Nam}},
  \bibinfo{author}{\bibfnamefont{W.~S.} \bibnamefont{Kolthammer}},
  \bibinfo{author}{\bibfnamefont{J.}~\bibnamefont{Huh}}, \bibnamefont{and}
  \bibinfo{author}{\bibfnamefont{I.~A.} \bibnamefont{Walmsley}},
  \bibinfo{journal}{arXiv:1710.08655}  (\bibinfo{year}{2017}).

\bibitem[{\citenamefont{Sparrow et~al.}(2018)\citenamefont{Sparrow,
  Mart{\'\i}n-L{\'o}pez, Maraviglia, Neville, Harrold, Carolan, Joglekar,
  Hashimoto, Matsuda, O’Brien et~al.}}]{sparrow2018simulating}
\bibinfo{author}{\bibfnamefont{C.}~\bibnamefont{Sparrow}},
  \bibinfo{author}{\bibfnamefont{E.}~\bibnamefont{Mart{\'\i}n-L{\'o}pez}},
  \bibinfo{author}{\bibfnamefont{N.}~\bibnamefont{Maraviglia}},
  \bibinfo{author}{\bibfnamefont{A.}~\bibnamefont{Neville}},
  \bibinfo{author}{\bibfnamefont{C.}~\bibnamefont{Harrold}},
  \bibinfo{author}{\bibfnamefont{J.}~\bibnamefont{Carolan}},
  \bibinfo{author}{\bibfnamefont{Y.~N.} \bibnamefont{Joglekar}},
  \bibinfo{author}{\bibfnamefont{T.}~\bibnamefont{Hashimoto}},
  \bibinfo{author}{\bibfnamefont{N.}~\bibnamefont{Matsuda}},
  \bibinfo{author}{\bibfnamefont{J.~L.} \bibnamefont{O’Brien}},
  \bibnamefont{et~al.}, \bibinfo{journal}{Nature}
  \textbf{\bibinfo{volume}{557}}, \bibinfo{pages}{660} (\bibinfo{year}{2018}).

\bibitem[{\citenamefont{Arrazola and Bromley}(2018)}]{arrazola2018using}
\bibinfo{author}{\bibfnamefont{J.~M.} \bibnamefont{Arrazola}} \bibnamefont{and}
  \bibinfo{author}{\bibfnamefont{T.~R.} \bibnamefont{Bromley}},
  \bibinfo{journal}{Phys. Rev. Lett.} \textbf{\bibinfo{volume}{121}},
  \bibinfo{pages}{030503} (\bibinfo{year}{2018}).

\bibitem[{\citenamefont{Arrazola et~al.}(2018)\citenamefont{Arrazola, Bromley,
  and Rebentrost}}]{arrazola2018quantum}
\bibinfo{author}{\bibfnamefont{J.~M.} \bibnamefont{Arrazola}},
  \bibinfo{author}{\bibfnamefont{T.~R.} \bibnamefont{Bromley}},
  \bibnamefont{and}
  \bibinfo{author}{\bibfnamefont{P.}~\bibnamefont{Rebentrost}},
  \bibinfo{journal}{Phys. Rev. A} \textbf{\bibinfo{volume}{98}},
  \bibinfo{pages}{012322} (\bibinfo{year}{2018}).

\bibitem[{\citenamefont{Br{\'a}dler et~al.}(2018)\citenamefont{Br{\'a}dler,
  Dallaire-Demers, Rebentrost, Su, and Weedbrook}}]{bradler2018gaussian}
\bibinfo{author}{\bibfnamefont{K.}~\bibnamefont{Br{\'a}dler}},
  \bibinfo{author}{\bibfnamefont{P.-L.} \bibnamefont{Dallaire-Demers}},
  \bibinfo{author}{\bibfnamefont{P.}~\bibnamefont{Rebentrost}},
  \bibinfo{author}{\bibfnamefont{D.}~\bibnamefont{Su}}, \bibnamefont{and}
  \bibinfo{author}{\bibfnamefont{C.}~\bibnamefont{Weedbrook}},
  \bibinfo{journal}{Phys. Rev. A} \textbf{\bibinfo{volume}{98}},
  \bibinfo{pages}{032310} (\bibinfo{year}{2018}).

\bibitem[{\citenamefont{Yoshikawa et~al.}(2016)\citenamefont{Yoshikawa,
  Yokoyama, Kaji, Sornphiphatphong, Shiozawa, Makino, and
  Furusawa}}]{yoshikawa2016invited}
\bibinfo{author}{\bibfnamefont{J.-i.} \bibnamefont{Yoshikawa}},
  \bibinfo{author}{\bibfnamefont{S.}~\bibnamefont{Yokoyama}},
  \bibinfo{author}{\bibfnamefont{T.}~\bibnamefont{Kaji}},
  \bibinfo{author}{\bibfnamefont{C.}~\bibnamefont{Sornphiphatphong}},
  \bibinfo{author}{\bibfnamefont{Y.}~\bibnamefont{Shiozawa}},
  \bibinfo{author}{\bibfnamefont{K.}~\bibnamefont{Makino}}, \bibnamefont{and}
  \bibinfo{author}{\bibfnamefont{A.}~\bibnamefont{Furusawa}},
  \bibinfo{journal}{APL Photonics} \textbf{\bibinfo{volume}{1}},
  \bibinfo{pages}{060801} (\bibinfo{year}{2016}).

\bibitem[{\citenamefont{{Vernon} et~al.}(2018)\citenamefont{{Vernon},
  {Quesada}, {Liscidini}, {Morrison}, {Menotti}, {Tan}, and
  {Sipe}}}]{vernon2018scalable}
\bibinfo{author}{\bibfnamefont{Z.}~\bibnamefont{{Vernon}}},
  \bibinfo{author}{\bibfnamefont{N.}~\bibnamefont{{Quesada}}},
  \bibinfo{author}{\bibfnamefont{M.}~\bibnamefont{{Liscidini}}},
  \bibinfo{author}{\bibfnamefont{B.}~\bibnamefont{{Morrison}}},
  \bibinfo{author}{\bibfnamefont{M.}~\bibnamefont{{Menotti}}},
  \bibinfo{author}{\bibfnamefont{K.}~\bibnamefont{{Tan}}}, \bibnamefont{and}
  \bibinfo{author}{\bibfnamefont{J.~E.} \bibnamefont{{Sipe}}},
  \bibinfo{journal}{arXiv:1807.00044}  (\bibinfo{year}{2018}).

\bibitem[{\citenamefont{Walschaers
  et~al.}(2017{\natexlab{a}})\citenamefont{Walschaers, Fabre, Parigi, and
  Treps}}]{walschaers2017entanglement}
\bibinfo{author}{\bibfnamefont{M.}~\bibnamefont{Walschaers}},
  \bibinfo{author}{\bibfnamefont{C.}~\bibnamefont{Fabre}},
  \bibinfo{author}{\bibfnamefont{V.}~\bibnamefont{Parigi}}, \bibnamefont{and}
  \bibinfo{author}{\bibfnamefont{N.}~\bibnamefont{Treps}},
  \bibinfo{journal}{Physical Review Letters} \textbf{\bibinfo{volume}{119}},
  \bibinfo{pages}{183601} (\bibinfo{year}{2017}{\natexlab{a}}).

\bibitem[{\citenamefont{Walschaers
  et~al.}(2017{\natexlab{b}})\citenamefont{Walschaers, Fabre, Parigi, and
  Treps}}]{walschaers2017statistical}
\bibinfo{author}{\bibfnamefont{M.}~\bibnamefont{Walschaers}},
  \bibinfo{author}{\bibfnamefont{C.}~\bibnamefont{Fabre}},
  \bibinfo{author}{\bibfnamefont{V.}~\bibnamefont{Parigi}}, \bibnamefont{and}
  \bibinfo{author}{\bibfnamefont{N.}~\bibnamefont{Treps}},
  \bibinfo{journal}{Physical Review A} \textbf{\bibinfo{volume}{96}},
  \bibinfo{pages}{053835} (\bibinfo{year}{2017}{\natexlab{b}}).

\bibitem[{\citenamefont{Chabaud et~al.}(2017)\citenamefont{Chabaud, Douce,
  Markham, Van~Loock, Kashefi, and Ferrini}}]{chabaud2017continuous}
\bibinfo{author}{\bibfnamefont{U.}~\bibnamefont{Chabaud}},
  \bibinfo{author}{\bibfnamefont{T.}~\bibnamefont{Douce}},
  \bibinfo{author}{\bibfnamefont{D.}~\bibnamefont{Markham}},
  \bibinfo{author}{\bibfnamefont{P.}~\bibnamefont{Van~Loock}},
  \bibinfo{author}{\bibfnamefont{E.}~\bibnamefont{Kashefi}}, \bibnamefont{and}
  \bibinfo{author}{\bibfnamefont{G.}~\bibnamefont{Ferrini}},
  \bibinfo{journal}{Physical Review A} \textbf{\bibinfo{volume}{96}},
  \bibinfo{pages}{062307} (\bibinfo{year}{2017}).

\bibitem[{\citenamefont{Lund et~al.}(2017)\citenamefont{Lund, Rahimi-Keshari,
  and Ralph}}]{lund2017exact}
\bibinfo{author}{\bibfnamefont{A.}~\bibnamefont{Lund}},
  \bibinfo{author}{\bibfnamefont{S.}~\bibnamefont{Rahimi-Keshari}},
  \bibnamefont{and} \bibinfo{author}{\bibfnamefont{T.}~\bibnamefont{Ralph}},
  \bibinfo{journal}{Physical Review A} \textbf{\bibinfo{volume}{96}},
  \bibinfo{pages}{022301} (\bibinfo{year}{2017}).

\bibitem[{\citenamefont{Hadfield}(2009)}]{hadfield2009single}
\bibinfo{author}{\bibfnamefont{R.~H.} \bibnamefont{Hadfield}},
  \bibinfo{journal}{Nature Photonics} \textbf{\bibinfo{volume}{3}},
  \bibinfo{pages}{696} (\bibinfo{year}{2009}).

\bibitem[{\citenamefont{Gupt et~al.}(2018)\citenamefont{Gupt, Arrazola,
  Quesada, and Bromley}}]{gupt2018classical}
\bibinfo{author}{\bibfnamefont{B.}~\bibnamefont{Gupt}},
  \bibinfo{author}{\bibfnamefont{J.~M.} \bibnamefont{Arrazola}},
  \bibinfo{author}{\bibfnamefont{N.}~\bibnamefont{Quesada}}, \bibnamefont{and}
  \bibinfo{author}{\bibfnamefont{T.~R.} \bibnamefont{Bromley}},
  \bibinfo{journal}{arXiv preprint arXiv:1810.00900}  (\bibinfo{year}{2018}).

\bibitem[{\citenamefont{Weedbrook et~al.}(2012)\citenamefont{Weedbrook,
  Pirandola, Garc{\'\i}a-Patr{\'o}n, Cerf, Ralph, Shapiro, and
  Lloyd}}]{weedbrook2012gaussian}
\bibinfo{author}{\bibfnamefont{C.}~\bibnamefont{Weedbrook}},
  \bibinfo{author}{\bibfnamefont{S.}~\bibnamefont{Pirandola}},
  \bibinfo{author}{\bibfnamefont{R.}~\bibnamefont{Garc{\'\i}a-Patr{\'o}n}},
  \bibinfo{author}{\bibfnamefont{N.~J.} \bibnamefont{Cerf}},
  \bibinfo{author}{\bibfnamefont{T.~C.} \bibnamefont{Ralph}},
  \bibinfo{author}{\bibfnamefont{J.~H.} \bibnamefont{Shapiro}},
  \bibnamefont{and} \bibinfo{author}{\bibfnamefont{S.}~\bibnamefont{Lloyd}},
  \bibinfo{journal}{Reviews of Modern Physics} \textbf{\bibinfo{volume}{84}},
  \bibinfo{pages}{621} (\bibinfo{year}{2012}).

\bibitem[{\citenamefont{Serafini}(2017)}]{serafini2017quantum}
\bibinfo{author}{\bibfnamefont{A.}~\bibnamefont{Serafini}},
  \emph{\bibinfo{title}{Quantum Continuous Variables: A Primer of Theoretical
  Methods}} (\bibinfo{publisher}{CRC Press}, \bibinfo{year}{2017}).

\bibitem[{\citenamefont{Bartlett and
  Sanders}(2002{\natexlab{a}})}]{bartlett2002universal}
\bibinfo{author}{\bibfnamefont{S.~D.} \bibnamefont{Bartlett}} \bibnamefont{and}
  \bibinfo{author}{\bibfnamefont{B.~C.} \bibnamefont{Sanders}},
  \bibinfo{journal}{Physical Review A} \textbf{\bibinfo{volume}{65}},
  \bibinfo{pages}{042304} (\bibinfo{year}{2002}{\natexlab{a}}).

\bibitem[{\citenamefont{Bartlett and
  Sanders}(2002{\natexlab{b}})}]{bartlett2002efficient}
\bibinfo{author}{\bibfnamefont{S.~D.} \bibnamefont{Bartlett}} \bibnamefont{and}
  \bibinfo{author}{\bibfnamefont{B.~C.} \bibnamefont{Sanders}},
  \bibinfo{journal}{Physical Review Letters} \textbf{\bibinfo{volume}{89}},
  \bibinfo{pages}{207903} (\bibinfo{year}{2002}{\natexlab{b}}).

\bibitem[{\citenamefont{Lita et~al.}(2008)\citenamefont{Lita, Miller, and
  Nam}}]{lita2008counting}
\bibinfo{author}{\bibfnamefont{A.~E.} \bibnamefont{Lita}},
  \bibinfo{author}{\bibfnamefont{A.~J.} \bibnamefont{Miller}},
  \bibnamefont{and} \bibinfo{author}{\bibfnamefont{S.~W.} \bibnamefont{Nam}},
  \bibinfo{journal}{Optics {E}xpress} \textbf{\bibinfo{volume}{16}},
  \bibinfo{pages}{3032} (\bibinfo{year}{2008}).

\bibitem[{\citenamefont{Marsili et~al.}(2013)\citenamefont{Marsili, Verma,
  Stern, Harrington, Lita, Gerrits, Vayshenker, Baek, Shaw, Mirin
  et~al.}}]{marsili2013detecting}
\bibinfo{author}{\bibfnamefont{F.}~\bibnamefont{Marsili}},
  \bibinfo{author}{\bibfnamefont{V.~B.} \bibnamefont{Verma}},
  \bibinfo{author}{\bibfnamefont{J.~A.} \bibnamefont{Stern}},
  \bibinfo{author}{\bibfnamefont{S.}~\bibnamefont{Harrington}},
  \bibinfo{author}{\bibfnamefont{A.~E.} \bibnamefont{Lita}},
  \bibinfo{author}{\bibfnamefont{T.}~\bibnamefont{Gerrits}},
  \bibinfo{author}{\bibfnamefont{I.}~\bibnamefont{Vayshenker}},
  \bibinfo{author}{\bibfnamefont{B.}~\bibnamefont{Baek}},
  \bibinfo{author}{\bibfnamefont{M.~D.} \bibnamefont{Shaw}},
  \bibinfo{author}{\bibfnamefont{R.~P.} \bibnamefont{Mirin}},
  \bibnamefont{et~al.}, \bibinfo{journal}{Nature Photonics}
  \textbf{\bibinfo{volume}{7}}, \bibinfo{pages}{210} (\bibinfo{year}{2013}).

\bibitem[{\citenamefont{Fiur{\'a}{\v{s}}ek
  et~al.}(2005)\citenamefont{Fiur{\'a}{\v{s}}ek, Garc{\'\i}a-Patr{\'o}n, and
  Cerf}}]{fiuravsek2005conditional}
\bibinfo{author}{\bibfnamefont{J.}~\bibnamefont{Fiur{\'a}{\v{s}}ek}},
  \bibinfo{author}{\bibfnamefont{R.}~\bibnamefont{Garc{\'\i}a-Patr{\'o}n}},
  \bibnamefont{and} \bibinfo{author}{\bibfnamefont{N.~J.} \bibnamefont{Cerf}},
  \bibinfo{journal}{Phys. Rev. A} \textbf{\bibinfo{volume}{72}},
  \bibinfo{pages}{033822} (\bibinfo{year}{2005}).

\bibitem[{\citenamefont{Bj{\"o}rklund et~al.}(2018)\citenamefont{Bj{\"o}rklund,
  Gupt, and Quesada}}]{bjorklund2018faster}
\bibinfo{author}{\bibfnamefont{A.}~\bibnamefont{Bj{\"o}rklund}},
  \bibinfo{author}{\bibfnamefont{B.}~\bibnamefont{Gupt}}, \bibnamefont{and}
  \bibinfo{author}{\bibfnamefont{N.}~\bibnamefont{Quesada}},
  \bibinfo{journal}{arXiv:1805.12498}  (\bibinfo{year}{2018}).

\bibitem[{\citenamefont{Neville et~al.}(2017)\citenamefont{Neville, Sparrow,
  Clifford, Johnston, Birchall, Montanaro, and Laing}}]{neville2017classical}
\bibinfo{author}{\bibfnamefont{A.}~\bibnamefont{Neville}},
  \bibinfo{author}{\bibfnamefont{C.}~\bibnamefont{Sparrow}},
  \bibinfo{author}{\bibfnamefont{R.}~\bibnamefont{Clifford}},
  \bibinfo{author}{\bibfnamefont{E.}~\bibnamefont{Johnston}},
  \bibinfo{author}{\bibfnamefont{P.~M.} \bibnamefont{Birchall}},
  \bibinfo{author}{\bibfnamefont{A.}~\bibnamefont{Montanaro}},
  \bibnamefont{and} \bibinfo{author}{\bibfnamefont{A.}~\bibnamefont{Laing}},
  \bibinfo{journal}{Nature Physics} \textbf{\bibinfo{volume}{13}},
  \bibinfo{pages}{1153} (\bibinfo{year}{2017}).

\bibitem[{\citenamefont{Clifford and Clifford}(2018)}]{clifford2018classical}
\bibinfo{author}{\bibfnamefont{P.}~\bibnamefont{Clifford}} \bibnamefont{and}
  \bibinfo{author}{\bibfnamefont{R.}~\bibnamefont{Clifford}}, in
  \emph{\bibinfo{booktitle}{Proceedings of the Twenty-Ninth Annual ACM-SIAM
  Symposium on Discrete Algorithms}} (\bibinfo{organization}{SIAM},
  \bibinfo{year}{2018}), pp. \bibinfo{pages}{146--155}.

\bibitem[{\citenamefont{Vere-Jones}(1988)}]{vere1988generalization}
\bibinfo{author}{\bibfnamefont{D.}~\bibnamefont{Vere-Jones}},
  \bibinfo{journal}{Linear Algebra and Its Applications}
  \textbf{\bibinfo{volume}{111}}, \bibinfo{pages}{119} (\bibinfo{year}{1988}).

\bibitem[{\citenamefont{Vere-Jones}(1997)}]{vere1997alpha}
\bibinfo{author}{\bibfnamefont{D.}~\bibnamefont{Vere-Jones}},
  \bibinfo{journal}{New Zealand J. Math} \textbf{\bibinfo{volume}{26}},
  \bibinfo{pages}{125} (\bibinfo{year}{1997}).

\bibitem[{\citenamefont{Crane}(2013)}]{crane2013some}
\bibinfo{author}{\bibfnamefont{H.}~\bibnamefont{Crane}},
  \bibinfo{journal}{Linear Algebra and Its Applications}
  \textbf{\bibinfo{volume}{439}}, \bibinfo{pages}{3445} (\bibinfo{year}{2013}).

\bibitem[{\citenamefont{Lu and Richards}(2001)}]{lu2001macmahon}
\bibinfo{author}{\bibfnamefont{I.-L.} \bibnamefont{Lu}} \bibnamefont{and}
  \bibinfo{author}{\bibfnamefont{D.~S.~P.} \bibnamefont{Richards}},
  \bibinfo{journal}{Advances in Applied Mathematics}
  \textbf{\bibinfo{volume}{27}}, \bibinfo{pages}{531} (\bibinfo{year}{2001}).

\bibitem[{\citenamefont{Konvalinka and Pak}(2006)}]{konvalinka2006non}
\bibinfo{author}{\bibfnamefont{M.}~\bibnamefont{Konvalinka}} \bibnamefont{and}
  \bibinfo{author}{\bibfnamefont{I.}~\bibnamefont{Pak}},
  \bibinfo{journal}{arXiv preprint math/0607737}  (\bibinfo{year}{2006}).

\bibitem[{\citenamefont{Owen}(2013)}]{mcbook}
\bibinfo{author}{\bibfnamefont{A.~B.} \bibnamefont{Owen}},
  \emph{\bibinfo{title}{Monte Carlo theory, methods and examples}}
  (\bibinfo{year}{2013}).

\end{thebibliography}

\appendix
\section{Click probabilities with threshold detectors}\label{derivation}
In this section we calculate the probability of a click pattern event when a Gaussian state is measured with \thr \ detectors. 
We focus on the case where the mean displacements are all zero, $\bm{\bar \alpha}_i = \bar \r_i = 0 \ \forall_i$.
We write the Gaussian $Q$ function with covariance matrix matrix $\Q$ as
\eq{\label{Eq:Q(alpha)}
	Q(\bm{\alpha}) = \frac{\braket{\bm{\alpha} | \rho | \bm{\alpha}}}{\pi^\ell} = \frac{  \exp\left(-\frac{1}{2}\bm{\alpha}  \Q^{-1}   \bm{\alpha}^\dagger  \right)}{\pi^\ell \sqrt{\det(\Q)}}.
}	
It can be shown that the matrix $\bm{D} \equiv \Q^{-1}$ always has the same block structure as $\Q$, {given by}
\eq{
	\bm{D} = 
	\left[
	\begin{array}{c c}
		\bm{K} & \bm{L}^* \\
		\bm{L} & \bm{K}^* \\
	\end{array}
	\right]>0,\label{Eq:Dblockm}
}
where $\bm{K}=\bm{K}^\dagger$ is Hermitian and $\bm{L} = \bm{L}^T$ is symmetric.
{The $P$ functions {of the POVM elements} 
	$\pr{0}$ and $\mathbb{\hat I} - \pr{0}$ in mode $n$ can be written as}
\eq{\label{vac0P}
	P^{(n)}_0(\alpha_n) = \delta^2(\alpha_n), \
	P^{(n)}_1(\alpha_n) = 
	\frac{1}{\pi} -P^{(n)}_0(\alpha_n).
}
{Suppose an $\ell$-mode Gaussian state is measured with threshold detectors and $N$ clicks are observed in the modes indexed by $S = \{ i_1,i_2,\ldots,i_{N}\}$. The probability of this event is given by}
\eq{\label{ps}
	p(S) = \pi^\ell \int  \prod_{i \in S} d^2 \alpha_i P^{(i)}_{1}(\alpha_i) 
	\prod_{k \notin S} d^2 \alpha_k P^{(k)}_{0}(\alpha_k) 
	Q(\bm{\alpha}).
}	
Whenever there is no click in mode $k$, we are forced to {set} $\alpha_k=\alpha^*_k=0$ in Eq. (\ref{ps}). Thus we can just make a matrix $\bm{D}_{(S)}$ with {$2N$} rows and columns obtained from $\bm{D}$ by keeping only the rows and columns {$\{i_1,\ldots,i_{N},i_1+\ell,\ldots,i_{N}+\ell \}$} associated with the modes where no photon was detected. The matrix $\bm{D}_{(S)}$ has the same block structure of $\bm{D}$ in Eq. \eqref{Eq:Dblockm}, 
\eq{
	\bm{D}_{(S)} &=\left[
	\begin{array}{c c}
		\bm{K}_{(s)} & \bm{L}_{(s)}^* \\
		\bm{L}_{(s)} & \bm{K}_{(s)}^* \\
	\end{array}
	\right],
}
where the (lowercase) label $s$ is used to indicate {which} rows and columns {$\{i_1,\ldots,i_{N}\}$} have been kept from the matrices $\bm{K},\bm{L}$.
To simplify notation we now use the dummy integration variables $\bm{\beta} = (\beta_1,\ldots \beta_m,\beta_1^*,\ldots,\beta_N^*)^T$ for the probability in Eq. (\ref{ps}). {Employing Eq. \eqref{Eq:Q(alpha)} we obtain}
\eq{
	p(S) =& \frac{1}{\sqrt{\text{det}(\Q)}} \int \prod_{k=1}^N d^2 \beta_k P_{1}(\beta_{k}) \exp\left(-\frac{1}{2}\bm{\beta} \bm{D}_{(S)} \bm{\beta}^\dagger \right) \nonumber \\
	=&\frac{1}{\pi^N \sqrt{\text{det}(\Q)}}\int \prod_{k=1}^N d^2 \beta_k \left(1 - \pi \delta(\beta_k)\delta(\beta^*_k) \right) \nn \\
	& \quad \times   \exp\left(-\frac{1}{2}\bm{\beta} \bm{D}_{(S)} \bm{\beta}^\dagger  \right). 
}
Now we need to rewrite the product terms $1-\pi \delta(\beta_k)\delta(\beta^*_k)$. To this end we use the following identity:
\eq{
	\prod_{k=1}^N (1-x_k) = \sum_{Z \in P([N])} (-1)^{|Z|} \prod_{i=1}^{|Z|} x_{Z_i},
}
where {$[m]$ denotes the set of integers $\{1,2,\ldots,m \}$, $P(A)$ is the power set (the set of all subsets) of $A$, and $|A|$ indicates the cardinality of $A$.} 
For example, $[2]=\{1,2\}$, $P([2]) = \left\{ \{\}, \{1\}, \{ 2\},\{1,2\} \right\}$, and if we take $Z = \{1,2\}=\{Z_1,Z_2\}$, then  $|Z|=2$ and $\prod_{i=1}^{|Z|} x_{Z_i} = x_{Z_1} x_{Z_2}=x_1 x_2$.
With this simplified notation we can write
\eq{
	\prod_{k=1}^N \left(1 - \pi \delta(\beta_k)\delta(\beta^*_k) \right) = \sum_{Z \in P([N])} (-\pi)^{|Z|} \prod_{i=1}^{|Z|} \delta(\beta_{Z_i}) \delta(\beta^*_{Z_i}),
}
and the click probability becomes
\eq{
	&p(S) = \frac{1}{\pi^N  \sqrt{\text{det}(\Q)}} \int \prod_{k=1}^N d^2 \beta_k \\
	&\times \sum_{Z \in P([N])} (-\pi)^{|Z|} \prod_{i=1}^{|Z|} \delta(\beta_{Z_i}) \delta(\beta^*_{Z_i}) \exp\left(-\frac{1}{2}\bm{ \beta } \bm{D}_{(S)} \bm{ \beta}^\dagger  \right)\nonumber .
}
As before, each time we have a delta function in the variables $\beta_{i_k}, \beta_{i_k}^*$ we have to remove the columns $i_k$ and $i_k+l$ and whatever is left is a Gaussian integral.
For each element $Z=\{Z_1,\ldots,Z_{|Z|}\} \in P([N])$, we again use the notation $\bm{F}_{(Z)}$ for the $2|Z| \times 2|Z|$ matrix obtained from $\bm{F} \in \mathbb{C}^{2 N \times 2N}$ by keeping columns and rows $Z_1,\ldots,Z_{|Z|}$  and $Z_1+N,\ldots,Z_{|Z|}+N$. With this, we can write
\eq{
	p(S) &= \frac{1}{\pi^m  \sqrt{\text{det}(\Q)}} \sum_{Z \in P([N])}  \frac{(-\pi)^{|Z|} (\pi)^{N-|Z|}}{\sqrt{\det((\bm{D}_{(S)} )_{(Z)})}} \nn \\
	&= \frac{\text{Tor}(\one-\bm{D}_{(S)})}{\sqrt{\det(\Q)}},
}
where in the last line we introduced the Torontonian of the matrix $\bm{D}_{(S)}$. For any matrix $\bm{A} \in \mathbb{C}^{2N \times 2N}$ that can be written as in Eq. \eqref{Eq:Dblockm}, we define its Torontonian as
\eq{
	{\text{Tor}(\bm{A}) = \sum_{Z \in P([N])} (-1)^{|Z|}  \frac{1}{\sqrt{\det(\one-\bm{A}_{(Z)})}}},
}
which is precisely the equation used to give the probability of detection in Eq. (\ref{Eq:Torontonian}) of the main text.
\section{Hafnians and Gaussian Boson Sampling}\label{app2}
The Hafnian of a $2 \ell \times 2 \ell$ {symmetric complex matrix} is defined as
\eq{
	\text{Haf}(\bm{A}) = \sum_{\mu \in \text{PMP}} \prod_{j=1}^\ell A_{\mu(2j-1),\mu(2j)},
}
where PMP stands for the set of perfect matching permutations.
As defined, it takes  $(2 \ell -1)!! = 1 \times 3 \times 5 \times \ldots \times (2 \ell -1)$ operations to calculate the Hafnian of $\bm{A}$.
In Ref. \cite{bjorklund2018faster} the following formula for the Hafnian is derived:
\eq{\label{bjorklund}
	\text{Haf}(\bm{A}) = \sum_{Z \in P([\ell])} (-1)^{ |Z|} f\left((\bm{A} \bm{X})_{(Z)}\right),
}
where the matrix $\bm{X}$ is defined as
\eq{
	\bm{X}= \bm{X}^T=\bm{X}^{-1}= \begin{bmatrix}
		\bm{0} & \one \\
		\one & \bm{0}
	\end{bmatrix}.	
}
The function $f(\bm{C})$ takes a matrix  $\bm{C} $ and returns the coefficient of $z^\ell$ in the following polynomial:
\eq{\label{poly}
	p_{\ell}(\eta \bm{C}) = \sum_{j=1}^{\ell} \frac{1}{j!}\left(\sum_{k=1}^{\ell} \frac{\tr(\bm{C}^k)}{2k}\eta^k \right)^j.
}
This coefficient can be found by taking derivatives, i.e.,
\eq{\label{gen_func}
	f(\bm{C}) = \frac{1}{\ell!} \left. \frac{d^\ell}{d\eta^\ell} p_\ell(\eta \bm{C}) \right|_{\eta=0}.
}
{The function $p_\ell(\eta\bm{C})$ requires only the eigenvalues of the matrix $\bm{C}$, since it considers just traces of powers of $\bm{C}$, which can be calculated explicitly in terms of the eigenvalues}.
Note that the formula in Eq. (\ref{bjorklund}) is significantly faster than the naive definition of the Hafnian since it requires {a summation of $\sim |P([\ell]){\tiny }| = 2^{\ell}$ terms}. Note that, instead of considering $p_\ell(\eta \bm{C})$ in Eq. (\ref{gen_func}), one can consider $p_{\ell'}(\eta \bm{C})$ for any $\ell'>\ell$. This will only add polynomials of degree $\ell' >\ell$ which will not change the value of $f(\bm{C})$. In particular, one can let $\ell' \to \infty$. This will become important in the next section when we link the Hafnian and the Torontonian.

For Gaussian Boson Sampling (GBS), we need the Hafnian of the symmetric $\bm{X} \bm{O}$ where,
\eq{\label{Bcov}
	\bm{O} =  \one - \bm{D}.
}
In GBS the probability of an event is given by 
\eq{
	\text{Haf}(\bm{X} \bm{O})&=\text{Haf}(\bm{X} \bm{X} \bm{O} \bm{X}) =  \text{Haf}\left(\bm{X} \bm{X} \left( \one - \bm{D} \right)  \bm{X}  \right)\nn \\
	&= \sum_{Z \in P([\ell])} (-1)^{ |Z|}  f\left(  \left( \left( \one- \bm{D} \right)\bm{X}  \bm{X} \right)_{(Z)}\right)\nn \\
	&=\sum_{Z \in P([\ell])} (-1)^{ |Z|}  f\left(\one - \bm{D}_{(Z)}\right)\nn \\
	&=\sum_{Z \in P([\ell])} (-1)^{ |Z|}  f\left( \bm{O}_{(Z)}\right) 
}
In the first line we used the fact that the Hafnian of a matrix whose rows and columns have been permuted is equal to the Hafnian of the unpermuted matrix. In the second line we used the fact that $\bm{X}$ is Hermitian and its own inverse.\\
The last formula is rather interesting because it makes explicit that even if the covariance matrix corresponds to a mixed state, \emph{i.e.}, if  $\mathbf{W} \neq \one_{\ell}$, the Hafnian of the symmetric matrix $\bm{X O}$ is always a real number since $ \one - \bm{D}_{(Z)}$ is also a Hermitian matrix and thus the eigenvalues of all its principal submatrices are real.

\section{Connecting the Torontonian and the Hafnian}\label{app3}
The Hafnian of the matrix $\bm{X O}$ gives the probabilities of a certain click pattern in a photon-number resolving (PNR) detector. If instead we used threshold detectors, the probability of an event would be proportional to the Torontonian of $\bm{O}$ (see Eq. (\ref{Bcov})):
\eq{\label{torontonian_formula}
	\text{Tor}(\bm{O}) = 	\sum_{Z \in P([\ell])} (-1)^{ |Z|} g\left(\bm{O}_{(Z)}\right),
}
where now we have 
\eq{
	g(\bm{C}) = \frac{1}{\sqrt{\text{det}(\one-\bm{C})}}.
}
Like {the function $f$ introduced in Eq. (\ref{bjorklund}),} the function $g$ only depends on the eigenvalues of $\bm{C}$. {Indeed, note the strong similarities between the definition of the Torontonian in Eq. (\ref{torontonian_formula}) and the Hafnian formula in Eq. (\ref{bjorklund}).}

We can make this suggestive connection more explicit. Specifically, we can write the Hafnian in terms of the Torontonian as 
\eq{
	\text{Haf}(\bm{X} \bm{O}) &=  \frac{1}{\ell!} \left. \frac{d^\ell}{d\eta^\ell} \left( 
	\sum_{Z \in P([\ell])} (-1)^{ |Z|} g\left(\eta \bm{O}_{(Z)}\right)	\right)	 \right|_{\eta=0}\nn \\
	&=\frac{1}{\ell!} \left. \frac{d^\ell}{d\eta^\ell} \text{Tor}(\eta \bm{O}) \right|_{\eta=0}.
}
{To see this, we extend the limits of the sums in Eq. (\ref{poly}) to infinity, since this does not affect the coefficient in front of $z^\ell$. We can therefore redefine
	\begin{align}
	\label{polyn}
	p_{\ell}(\eta\bm{C}) & \rightarrow \sum_{j=1}^{\infty} \frac{1}{j!}\left(\sum_{k=1}^{\infty} \frac{\tr(\bm{C}^k)}{2k}\eta^k \right)^j \\
	& = \exp\left(\sum_{k=1}^{\infty} \frac{\tr(\bm{C}^k)}{2k}\eta^k \right).
	\end{align}
	From this form, we can recognize the Mercator series for the logarithm:
	\eq{
		-\frac{1}{2} \log {\rm det}\, (\one- \eta\bm{C}) &= -\frac{1}{2} {\rm Tr} (\log{(\one-\eta\bm{C})}) \nn \\
		&=\sum_{k=1}^\infty \frac{{\rm Tr} (\bm{C}^k)}{2k}\eta^k.
	}
	By taking the exponential on both sides, we get
	\eq{
		\exp&\left(-\frac{1}{2} \log {\rm det}\, (\one- \eta\bm{C}) \right)=
		\frac{1}{\sqrt{\text{det}(\one- \eta\bm{C})}} \nn \\
		&=\sum_{j=0}^{\infty} \frac{1}{j!} \left(\sum_{k=1}^\infty \frac{{\rm Tr} (\bm{C})^k}{2k}\eta^k  \right)^j.
	}
	We conclude that the generating function $p_\ell(\eta\bm{C})$ from Eqs. (\ref{poly})-(\ref{gen_func}) can be replaced by
	\begin{equation}
	p_\ell(\eta\bm{C}) = \frac{1}{\sqrt{\text{det}(\one- \eta\bm{C})}},
	\end{equation}
	which establishes the connection of Eq. (\ref{hafmgf}) between Hafnians and Torontonians.
	
	Finally, we note that the function $[\text{det}(\one- \eta\bm{C})]^{-\alpha}$ has been explored in previous works in the literature. Depending on the choice of $\alpha$, it can be seen as a generating function for determinants, permanents, and generalizations of permanents called $\alpha$-permanents \cite{vere1988generalization, vere1997alpha, crane2013some}. It also appears in generalizations of the MacMahon master theorem \cite{lu2001macmahon, konvalinka2006non}.
}

\section{Gaussian measurements in states that are linear combinations of Gaussians}\label{app4}
In this section we investigate how to generate samples of homodyne and heterodyne measurements applied to states that are linear combinations of Gaussian states,
\eq{
	\p_{\ell} = \sum_{k=1}^N 	a_k \p_{\ell,k}(\bm{V}_k,\bm{\bar r}_k).
}
To simulate a measurement on mode $n$ we first find the marginal state of this mode, which is again a linear combination 
\eq{
	\sigma  = \sum_{k=1}^N a_k  \sigma_{k}(\bm{V}_{n,k},\bm{\bar r}_{n,k}),
}
where $\sigma_k(\bm{V}_{n,k},\bm{\bar r}_{n,k})$ is a single mode Gaussian state with covariance matrix $\bm{V}_{n,k}$ and vector of means $\bm{\bar r}_{n,k}$. Consider a Gaussian measurement with {POVM $\{\Pi(\bm{W},\bm{r}_m)\}_{\bm{r}_m}$}. To obtain the probability density function of a Gaussian measurement on mode $n$ (assumed without loss of generality to be the last one, so $n=\ell$), we use Born's rule to write
\eq{
	p(\bm{r}_m) &= \tr\left[ \hat \Pi(\bm{W},\bm{r}_m)  \sigma \right] = 
	\sum_{k=1}^N 	a_k q_k (\bm{r}_m),\\
	q_k (\bm{r}_m) &= \tr\left[  \hat \Pi(\bm{W},\bm{r}_m)  \sigma_{k}(\bm{V}_{n,k},\bm{\bar r}_{n,k}) \right],
}
where $\bm{r}_m$ is a two-dimensional vector. Note that each of the overlaps $\tr\left[  \hat \Pi(\bm{W},\bm{r}_m) \sigma_{k}(\bm{V}_{n,k},\bm{\bar r}_{n,k}) \right]$ is a Gaussian function in $\bm{r}_m$ and can be calculated in closed form by writing the Wigner function of each density matrix or POVM element and then doing Gaussian integrals in two-dimensional phase space.
The covariance matrix for homodyne measurements is  $\bm{W}_{\text{hom}} = \left[ \begin{smallmatrix}
1/s^2 & 0 \\
0 & s^2  
\end{smallmatrix} \right]$ with $s \gg 1$ and for heterodyne measurements $\bm{W}_{\text{het}} =  \left[ \begin{smallmatrix}
1 & 0 \\
0 & 1  
\end{smallmatrix} \right]$.

We now need to sample from this two-dimensional distribution, for which many methods are readily available (c.f. Chap. 5 of Ref. \cite{mcbook}). Also note that since the {probability density function} is a sum of Gaussians, we can easily obtain analytical expressions for the marginal {density functions} and cumulative distribution functions. Once a value $\bm{\tilde r}_m$ has been sampled with probability $p(\bm{\tilde r}_m)$, we can propagate the backaction by the following recipe:
\eq{
	\p_{\ell-1} &= \frac{\tr_{n=\ell}	\left[ \hat \Pi(\bm{W},\bm{\tilde r}_m)  \p_{\ell} \right] }{p(\bm{\tilde r}_m) } \\
	&= \frac{1}{p(\bm{\tilde r}_m)}  \sum_{k=1}^N 	a_k  \tr_{n=\ell}	\left[ \hat \Pi(\bm{W},\bm{\tilde r}_m)  \p_{\ell,k}(\bm{V}_k,\bm{\bar r}_k) \right].\nn
}
The covariance matrix, vector of means, and normalization of the unnormalized $(\ell-1)$-mode Gaussian state 
\eq{
	\tr_{n=\ell}	\left[ \hat \Pi(\bm{W},\bm{\tilde r}_m)  \p_{\ell,k}(\bm{V}_k,\bm{\bar r}_k) \right]=q_k(\bm{\tilde r}_m) \  \p_{\ell,k}\left(\bm{V}_{A,k}',\bm{\bar r}_{A,k}' \right)
}
are easily calculated by writing \cite{serafini2017quantum}
\eq{
	\V_k &= \begin{bmatrix}
		\noindent	\V_{A,k} & \bm{V}_{AB,k} \\
		\bm{V}_{AB,k}^T & \bm{V}_{B,k}
	\end{bmatrix}, \ \bm{\bar r}_k = \begin{bmatrix} \bm{\bar r}_{A,k} \\ \bm{\bar r}_{B,k} \end{bmatrix},\\
	\bm{V}_{A,k}' &= \bm{V}_{A,k} - \bm{V}_{AB,k} (\bm{V}_{B,k}+\bm{W} )^{-1} \bm{V}_{AB,k}^T,\\
	\bm{\bar r}_{A,k}'&=\bm{\bar r}_{A,k}+ \bm{V}_{AB,k} (\bm{V}_{B,k}+\bm{W})^{-1} (\bm{\tilde r}_m-\bm{\bar r}_{B,k}).
}
These results allows us to generate homodyne and heterodyne samples of the nongaussian states obtained by postselecting Gaussian states using \thr \ detectors. As discussed in Sec. \ref{sec:sampling} the exponential growth in the complexity of generating these samples is dictated only by the number of clicks necessary to generate the nongaussian state being sampled with heterodyne/homodyne measurements.

\section{Complexity of threshold Gaussian Boson Sampling}\label{app5}

In Gaussian Boson Sampling, let $p(S)$ be the probability of observing an output pattern $S$ when sampling from a state using PNRs, and let $p'(S)$ be the probability when using threshold detectors. Define the set of collision outputs $\mathcal{C}$ as the set of PNR outputs where two or more photons are observed in at least one mode. The probability of observing a collision when sampling from $p(S)$ is $\varepsilon := \sum_{S\in \mathcal{C}} p(S)$.  It holds that
\begin{align}
2\|p(S)-p'(S)\|_1&=\sum_S |p(S)-p'(S)|\nonumber\\
&= \sum_{S\in\Col}  |p(S)-p'(S)|+ \sum_{S\notin\Col}  |p(S)-p'(S)|\nonumber\\
&=\sum_{S\in\Col}  |p(S)|+ \sum_{S\notin\Col}  |p(S)-p'(S)|\nonumber\\
&=\varepsilon+ \sum_{S\notin\Col}  |p(S)-p'(S)|,\label{Eq: eps+p(S)}
\end{align}
where we have used the fact that $p'(S)=0$ for all $S\in\mathcal{C}$. Furthermore, let $\mathcal{C}_S$ be the set of PNR outputs where photons are observed only in the modes corresponding to the elements of $S$ and there is at least one mode where two or more photons are detected. Define the mapping $T$ such that {$T(S')=S\in\mathcal{C}$ for any $S'\in \Col_S$}, i.e., the mapping that takes collision outputs to outputs without any collisions. We then have that
\beq
p'(S)=p(S)+\sum_{S':T(S')=S}p(S'),
\eeq
which implies
\begin{align}
\sum_{S\notin\Col} |p(S)-p'(S)|&= \sum_{S\notin\Col} |p(S)-p(S)+\sum_{S':T(S')=S}p(S')|\nonumber\\
&=\sum_{S\notin\Col}\sum_{S':T(S')=S}p(S')\nonumber\\
&= \sum_{S'\in \Col}p(S')=\varepsilon,\label{Eq: eps}
\end{align}
where we have used the fact that $\Col = \bigcup_{S\notin\Col}\{S':T(S')=S\}$. From Eqs. \eqref{Eq: eps+p(S)} and \eqref{Eq: eps} we conclude that
\beq
\|p(S)-p'(S)\|_1=\varepsilon.
\eeq
Note that since $p(S)=\text{Haf}[\bm{X}\mathbf{O}_{(S)}]/\sqrt{\det{\Q}}$ and $p'(S)=\Tor[\mathbf{O}_{(S)}]/\sqrt{\det{\Q}}$, it also holds that
\beq
\| \text{Haf}[\bm{X}\mathbf{O}_{(S)}]-\Tor[\mathbf{O}_{(S)}]\|_1=\frac{\varepsilon}{\sqrt{\det{\Q}}}.
\eeq

\subsection{Collision probability}

It was proven in Ref. \citep{aaronson2011computational} that the probability of observing a collision when $2N$ identical photons interact in an $\ell$-mode linear interferometer satisfies
\beq
\mathbb{E}_\mathcal{U}[P(\text{collision})]<\frac{8N^2}{\ell},
\eeq 
where $\mathcal{U}$ is the unitary describing the interferometer and the expectation is taken over the Haar measure. For Gaussian Boson Sampling, the input photon number is not fixed but, since the linear interferometer commutes with the number operator, we can equivalently consider first performing a measurement of the total photon number and then applying the interferometer transformation. Denoting by $q(N)$ the probability of observing $N$ {total} photons, we have
\begin{align}
\mathbb{E}_\mathcal{U}[P(\text{collision})]&=\mathbb{E}_\mathcal{U}\left[\sum_{n=0}^\infty q(N)P(\text{collision}|N)\right]\nonumber\\
&= \sum_{n=0}^\infty q(N)\mathbb{E}_\mathcal{U}[P(\text{collision}|N)]\nonumber\\
&<\sum_{n=0}^\infty q(N)\frac{8N^2}{\ell}\nonumber\\
&=\frac{8}{\ell}\sum_{n=0}^\infty q(N) N^2\nonumber\\
&=\frac{8}{\ell}\mathbb{E}_q[N^2].
\end{align}

\end{document}